\documentclass{svjour2}                    % onecolumn
\smartqed  % flush right qed marks, e.g. at end of proof
\usepackage{graphicx}
\usepackage{amsmath}
\usepackage{amssymb}
\usepackage{mathptmx}      % use Times fonts if available on your TeX system
\usepackage{latexsym}
\def\Xint#1{\mathchoice
{\XXint\displaystyle\textstyle{#1}}%
{\XXint\textstyle\scriptstyle{#1}}%
{\XXint\scriptstyle\scriptscriptstyle{#1}}%
{\XXint\scriptscriptstyle\scriptscriptstyle{#1}}%
\!\int}
\def\XXint#1#2#3{{\setbox0=\hbox{$#1{#2#3}{\int}$}
\vcenter{\hbox{$#2#3$}}\kern-.5\wd0}}

\def\dashint{\Xint-}

\begin{document}

\title{Pauli matrices and 2D electron gas}
%\title{Boltzmann transport in the measuring instrument explains quantum correlation and Mach-Zehnder interferometry.}
%\subtitle{Paradox explained by instrument behavior and a note on meaning}
\author{J.F. Geurdes % etc
\thanks{\emph{Present address:} C. vd Lijnstraat 164 2593 NN, Den Haag, Netherlands}%
}                     % Do not remove
\institute{}
\date{Received: date / Accepted: date}
% Add name of the expert who has communicated your paper
%\communicated{name}
%
\maketitle
\begin{abstract}
In the present paper it will be argued that transport in a 2D electron gas can be implemented as 'local hidden instrument based' variables. With this concept of instrumentalism it is possible to explain the quantum correlation, the particle-wave duality and Wheeler's 'backward causation of a particle'. In the case of quantum correlation the spin measuring variant of the Einstein Podolsky and Rosen paradox is studied. In the case of particle-wave duality the system studied is single photon Mach-Zehnder (MZ) interferometry with a phase shift size $\delta$. The idea that the instruments more or less neutrally may show us the way to the particle will be replaced by the concept of laboratory equipment contributing in an unexpected way to the measurement.
\end{abstract}


\section{INTRODUCTION}
\label{sect:intro} 
Foundational quantum theory has posed many riddles to physics and one of the most important ones is the particle-wave duality. Another such big problem is causation. As is well-known Einstein (e.g. \cite{Eins31}) never fully agreed with the foundational nature of the quantum theory.  Many arguments were put against Einsteins position, i.e. \cite{Bell64} and \cite{CHSH}. The present author showed that doubt can be cast on the objections against Einsteins position by showing that a Monte Carlo type local hidden variables model can violate CHSH \cite{JFGOE}. Moreover, the present author and e.g. dr. Hans Salhofer independently from each other noted a strong resemblance between classical e.m. field theory and modern relativistic quantum theory \cite{JFG2}, \cite{JFG} and \cite{Sal}. In addition Rolin Armour demonstrated spin 1/2 fields for photons \cite{Arm}. The present concepts are in a way related to the previous notionsand resemble the ideas of Madelung \cite{Madel}. Madelung tried to give an interpretation of Schr{\"o}dinger's equation in terms of hydrodynamics.
 
The paper will argue for an instrumentalist interpretation of the quantum theory. Its claim is: the hidden variables of Einstein are, or reside in, the measuring instruments. An axiom is that only in ensembles particles have 'individuality'. This allows to employ Boltzmann theory of charged 'particles' in an instrument that demonstrates the activities of  a single particle like a photon. Moreover, acitivities of the observer change the set up of the experiment in a non neutral way. The setup can be expressed as 2D electron gas conductances inside the instrument. The concept of currents and conductances is applied in the form of an algebra to the Mach-Zehnder interferometer for a 'one photon-at-the-time' measurement. 

Elements of the interferometer, like mirrors, beamsplitters and phase shifter are represented in terms of changes to 2D electron gas conductances in the instrument. This provides what can be called a Mach-Zehnder (MZ) conductance algebra. With the use of this algebra Wheeler's backward causation paradox can be explained. The concept of this algebra implies that changes in the 'algebra of' instrument(s) explain the paradox of backward causation. Moreover, because Young's double slit interference experiment and MZ interferometry are conceptually the same, the MZ conductance algebra also explains the famous wave-particle dualism. The message is that crucial changes in measuremental set-up produces a different conductance algebra. 
  
In the following, three main chapters will be found. The first chapter is the derivation of the 2D conductances. The subsequent chapter is devoted to explaining the quantum correlation. The chapter following that is devoted to the Mach-Zehnder interferometer. The paper ends with a discussion section and remarks on meaningfulness of models.

\section{2D ELECTRON GAS AND PAULI MATRIX FORM CONDUCTANCES}
In the present chapter we first will introduce the 2D electron gas Boltzmann transport equation and derive two conductance tensors, $\sigma_x$ and $\sigma_y$, which have the form of Pauli matrices.  
%\subsection{ Boltzmann transport}
The Fermi-Dirac distribution of electrons in a two dimensional electron gas without electric and magnetic fields equals \cite{Barnes}
\begin{equation}\label{FermDir}
f_0(\mathbf{k})=\frac{1}{1+\exp\left[ \frac{\epsilon(\mathbf{k},t)-\mu}{k_B T}\right]}
\end{equation}
Here, $\mathbf{k}=(k_x,k_y)$ is the wave vector, $k_B$ is the  Boltzmann constant, $T$ the absolute temperature and $\mu$ the chemical potential. If an electromagnetic source is available the acceleration $\frac{d}{dt}\mathbf{v}(\mathbf{k})$ of the electrons is influenced by the Lorentz force.
\begin{equation}\label{Lorentz}
\frac{1}{\beta(\mathbf{k})}\frac{\hbar}{m^{*}}\frac{d \mathbf{k}}{dt}=\frac{d}{dt}\mathbf{v}(\mathbf{k})=-\frac{e}{m^{*}}\left(\mathbf{E}(\mathbf{k}) + \mathbf{v}(\mathbf{k}) \times \mathbf{B}(\mathbf{k})\right) 
\end{equation}
The acceleration is related to velocity in the wave vector $\frac{\hbar}{m^{*}}\dot{\mathbf{k}}$, with, $m^{*}$ the effective mass of the electron. The $\beta(\mathbf{k})$ is a coupling function between $\dot{\mathbf{k}}$, $\dot{\mathbf{v}}(\mathbf{k})$ and the Lorentz force. This coupling can be viewed upon as related to the medium in which the electron gas is confined. 

%\subsection{Current}
Because of the Lorentz force, the distribution $f=f(\mathbf{k},t)$ now differs from the one in equation (\ref{FermDir}). However, it is still assumed that $f(\mathbf{k},t)=f(\mathbf{k}+\dot{\mathbf{k}}dt,t+dt)$ with $\dot{\mathbf{k}}=\frac{d}{dt}{\mathbf{k}}$ etc. If the difference between $f(\mathbf{k},t)$ and $f_0(\mathbf{k})$ is denoted by $g(\mathbf{k},t)$ the current can be approximated by
\begin{equation}\label{curr}
\mathbf{j}(t)=-2\int \frac{d^2\mathbf{k}}{(2\pi)^2}\,e\mathbf{v}(\mathbf{k}) g(\mathbf{k},t)
\end{equation}  
The $2$ in front of the integral of equation (\ref{curr}) arises from the electron spin (considered phenomenologically). Note $\dot{\mathbf{k}}\neq \mathbf{0}$. The linear Boltzmann equation for $g(\mathbf{k},t)$ is then equal to
\begin{equation}\label{boltzmaneq}
\frac{g(\mathbf{k},t)}{ \tau(\mathbf{k})}= \left(\nabla_{\mathbf{k}} f_0(\mathbf{k})\right)\cdot \frac{e\beta(\mathbf{k})}{\hbar} \mathbf{E}(\mathbf{k})+\left(\nabla_{\mathbf{k}}g(\mathbf{k},t)\right)\cdot \frac{e\beta(\mathbf{k})}{\hbar} \mathbf{v}(\mathbf{k}) \times \mathbf{B}(\mathbf{k})
\end{equation}
\subsection{Semiclassical conductance}\label{sect:semco}
In a zero magnetic field or when $\nabla_{\mathbf{k}} g(\mathbf{k},t) \sim \mathbf{v}(\mathbf{k})+\mathbf{B}(\mathbf{k})$ the linear Boltzmann equation can be solved by
\begin{equation}\label{gee}
g(\mathbf{k},t)= \tau(\mathbf{k})\frac{e\beta(\mathbf{k})}{\hbar} \left(\nabla_{\mathbf{k}} f_0(\mathbf{k})\right)\cdot \mathbf{E}(\mathbf{k})
\end{equation}
Here $ \tau(\mathbf{k})$ is the relaxation time. In the general theory\cite{Barnes} 
%\begin{equation}\label{nabf0}
$\nabla_{\mathbf{k}} f_0 = \hbar \frac{\partial f_0}{\partial \epsilon(\mathbf{k},t)}\nabla_{\mathbf{k}} \epsilon(\mathbf{k},t)$
%\end{equation}
with
\begin{equation}\label{lim}
\lim_{T \rightarrow 0}   \frac{\partial f_0}{\partial \epsilon(\mathbf{k},t)} = - \delta\left(\epsilon-E_F \right)
\end{equation}
and $\nabla_{\mathbf{k}} \epsilon(\mathbf{k},t)=\alpha(\mathbf{k})\mathbf{v}(\mathbf{k})$,with $\alpha(\mathbf{k})$ in our theory a general coupling between energy gradient in $\mathbf{k}$ space and the velocity. Because of non-zero velocities, (\ref{lim}) holds approximately\cite{Barnes}. The $E_F$ is the Fermi energy. Given, $\mathbf{j}=\sigma  \mathbf{E}$ the conductance $2 \times 2$ matrix can be written
\begin{equation}\label{conduct}
\sigma = \frac{e^2}{2\hbar \pi^2} \int d^2 \mathbf{k}\,\delta\left(\epsilon(\mathbf{k},t)-E_F \right) \alpha(\mathbf{k})\beta(\mathbf{k})\tau(\mathbf{k})\mathbf{v}(\mathbf{k})\otimes \mathbf{v}(\mathbf{k})
\end{equation}
The tensor product $\otimes$ is: $(\mathbf{x} \otimes \mathbf{y})_{i,j}=x_i y_j $ for $i,j=1,2$ and $\mathbf{x}=(x_1,x_2)$ similarly $\mathbf{y}=(y_1,y_2)$. The scalar coupling functions $\alpha$ and $\beta$ are introduced in (\ref{conduct}).

\subsection{Pauli matrices}
\label{sect:Pauli}
Subsequently, the wave vector $\mathbf{k}$ is transformed with $k_x=k\cos(\varphi)$ and $k_y=k\sin(\varphi)$. The Jacobian is $k=\sqrt{k_x^2 +k_y^2}$. Hence
\begin{equation}\label{conduct2}
\sigma = \frac{e^2}{2\hbar \pi^2} \int kdk\,d\varphi\,\delta\left(\epsilon(k,\varphi,t)-E_F \right) \alpha(k,\varphi)\beta(k,\varphi) \tau(k,\varphi)\mathbf{v}(k,\varphi)\otimes \mathbf{v}(k,\varphi)
\end{equation}
Note that, because $\dot{\mathbf{k}}\neq 0$, we also see $\dot{k}\neq 0$ and/or $\dot{\varphi}\neq 0$. The next step is to observe that $\epsilon(k,\varphi,t)$ in (\ref{conduct2}) depends on the velocity $\mathbf{v}(k,\varphi)$ (see Barnes\cite{Barnes}). The subsequent transformation is to transform to $\epsilon=\epsilon(k,t)=\frac{1}{2}m^{*}\mathbf{v}^2(k,t)$ and only see radial velocities i.e. $v^2(k,t)=\mathbf{v}^2(k,t)$. The second transformation is $\phi=\varphi$. For completeness, $k=k(\epsilon,\phi,t)$ and $\varphi = \varphi(\epsilon,\phi,t)=\phi$. We do not use quadratic dispersion for $\epsilon$. 

\subsubsection{Remark on $\alpha$ and $\beta$}
If we for the sake of the completeness of the argument take a $\mathbf{k}=(k_x,k_y)$ dependence for $\epsilon$ and not only radial $k=\sqrt{k_x^2 + k_y^2}$ dependence, then we see that a time differential for $\epsilon$ is equal to
\begin{equation}\label{Dif1}
\frac{d \epsilon }{dt} = \left(\nabla_{\mathbf{k}} \epsilon\right)\cdot \dot{\mathbf{k}}+\frac{\partial \epsilon}{\partial t}=m^{*}\mathbf{v}\cdot \dot{\mathbf{v}}
\end{equation}
with $\dot{\mathbf{v}}=\frac{d}{dt}{\mathbf{v}}$ etc. If we then note that $\dot{\mathbf{v}}=\frac{\hbar}{m^{*}\beta}\dot{\mathbf{k}}$ and $\nabla_{\mathbf{k}} \epsilon=\alpha \, \mathbf{v}$ we have the consistency requirement for $\alpha$ and $\beta$. 
\begin{equation}\label{partDiff1}
\frac{\partial \epsilon}{\partial t}=m^{*}\left(1-\frac{\alpha \beta}{\hbar}\right) \mathbf{v}\cdot \dot{\mathbf{v}}
\end{equation}
Hence, when (\ref{partDiff1}) is unequal to zero then $\alpha \beta $ may be unequal to $\hbar$.
\subsubsection{The Jacobian and the reformulation}
The Jacobian for the transformation is 
\begin{eqnarray}\label{Jac}
J=
\left| \begin{array}{cc} \partial k / \partial \epsilon & \partial k / \partial \phi \\
\partial \varphi / \partial \epsilon & \partial \varphi / \partial \phi \end{array} \right|
\end{eqnarray}
With this particular transformation $\partial \varphi / \partial \epsilon = 0$ and $\partial \varphi / \partial \phi =1$. Let us take $\epsilon=\frac{1}{2}m^{*}\mathbf{v}^2(k,t)$. If we assume, $v_1(k,\varphi,t)=f(k,t)\cos(\varphi)$ and $v_2(k,\varphi,t)=f(k,t)\sin(\varphi)$ we can have (suppressing $t$ for ease of notation) $\mathbf{v}^2(k)=f^{\,2}(k)$ hence radial $\epsilon$ kinetic-like energy, while in the transformed $(\epsilon,\phi)$ system $\mathbf{v}^{T}=(v_1(\epsilon,\phi), v_2(\epsilon,\phi))$.
Hence, 
\begin{equation}\label{auxdif}
1=m^{*}||\mathbf{v}(k)|| \frac{\partial ||\mathbf{v}(k)||}{\partial k} \frac{\partial k}{\partial \epsilon}
\end{equation}
With $||.||$ the Euclidean norm like e.g. in  $||\mathbf{k}||=k=\sqrt{k_x^2 +k_y^2}$. Hence, the Jacobian in (\ref{Jac}) is $J=w(\epsilon)=\partial k/\partial \epsilon$ which, in $k$ and $\varphi$, looks like $\{m^{*}||\mathbf{v}(k)|| \frac{\partial ||\mathbf{v}(k)||}{\partial k} \}^{-1}$. Generally, $||\mathbf{v}(k,\varphi)||>0$, is assumed despite $T\rightarrow 0$ leading to a delta function for $-\partial f_0 / \partial \epsilon$. The conductance in (\ref{conduct2}) now can be transformed into
\begin{equation}\label{conduct3}
\sigma = \frac{e^2}{2\hbar \pi^2} \int_0^{\infty} d\epsilon\,\int_{-\pi}^{+\pi} d\phi\,\delta\left(\epsilon-E_F \right) \tau^{\prime}(\epsilon,\phi)\mathbf{v}(\epsilon,\phi)\otimes \mathbf{v}(\epsilon,\phi)
\end{equation}
with $ \tau^{\prime}(\epsilon,\phi)=k(\epsilon,\phi) \tau(\epsilon,\phi)w(\epsilon)\alpha(\epsilon,\phi)\beta(\epsilon,\phi)$. It is assumed that $\tau^{\prime}(\epsilon,\phi)$ can be either positive or negative. The $\phi$ positive or negative dependency resides in $\beta(\epsilon,\phi)$.  This is so because, $||\mathbf{v}||=f(k)$ entails that $\alpha=\alpha(k)$ while $\epsilon=\frac{1}{2}m^{*}\mathbf{v}^2(k)$ implies $k=k(\epsilon)$ and $w(\epsilon)$ is independent of $\phi$. Subsequent integration over $\epsilon$  (with $E_F >0$) results into
\begin{equation}\label{conduct4}
\sigma = \frac{e^2}{2\hbar \pi^2}\int_{-\pi}^{+\pi} d\phi\ \tau^{\prime}(E_F,\phi)\mathbf{v}(E_F,\phi)\otimes \mathbf{v}(E_F,\phi)
\end{equation}
%\subsubsection{Steps towards Pauli matrices}
The first step to the Pauli matrices for (\ref{conduct4}) is to transform the velocity vector to an 'associated' form: $\mathbf{u}(E_F,\phi)=\mathbf{v}(E_F,\phi)\sqrt{ \tau^{\prime}(E_F,\phi)}$. Note that even when $\mathbf{v}$ is independent of $\phi$ the $\mathbf{u}$ can vary with $\phi$ because of $\phi$ dependence in $\tau^{\prime}$. $\mathbf{u}$ is real or imaginary and $\tau^{\prime}(E_F,\phi)=k(E_F) \tau(E_F,\phi)w(E_F)\alpha(E_F)\beta(E_F,\phi)$, with real relaxation time $ \tau(E_F,\phi)$ and, real positive or negative, Jacobian $w(E_F)$. Let us specify $u_i(E_F,\phi)$ for $i=1,2$ and derive the first Pauli matrix.

\subsubsection{Pauli's $\sigma_x$ for conductance}\label{P1}\label{velvec} 
Suppose, the Heaviside $H$ is defined by, $H(x)=1$ for all $x\geq 0$ and $H(x)=0$ for all $x<0$. Moreover, suppose $\psi \in (0,\frac{\pi}{4})$. Then for $\phi \in (-\psi,\psi)$,
\begin{eqnarray}\label{Paul1}
\begin{array}{cc} \delta u_{x,1}=u_1(E_F,\phi)=\kappa(E_F)\sqrt{\frac{1}{2\psi}}\sqrt{\tan(\phi)}H(\phi+\psi)H(\psi-\phi)\\
\delta u_{x,2}=u_2(E_F,\phi)=\kappa(E_F)\sqrt{\frac{1}{2\psi}}\sqrt{\cot(\phi)}H(\phi+\psi)H(\psi-\phi) \end{array}
\end{eqnarray}
Note that in this definition, suppressing the $E_F$ dependence notation for the moment, we see $u_i^2(\phi) <0$ when $\phi <0$ and  $u_i^2(\phi) \geq 0$ when $\phi \geq 0$. The indication $\delta u_{x,i}$ refers to the choice of relatively small changes in associated velocity for the Pauli matrix $\sigma_x$. For a given $\phi >0$ and $\phi \in (0,\psi)$ we assume in approximation the  conservation of total kinetic energy for 'differential associated velocities' $\delta \mathbf{u}_{x}(E,\phi)=\mathbf{u}(E,\phi)$. I.e. the changes in kinetic energies when electron and hole are created occur 'balanced' in the $\mathbf{u}$. 
\begin{equation}\label{som0}
\frac{1}{2}m^{*}||\mathbf{u}(E,\phi)||^2 + \frac{1}{2}m^{*}||\mathbf{u}(E,-\phi)||^2=0
\end{equation}  
It is believed that electron escape from a crystal structure leaving behind a hole can be pictured in equation (\ref{Paul1}). Note that equation (\ref{som0}) is valid in $\mathbf{u}$ not in $\mathbf{v}$ terms.
Integrating for $u^2_1$ for instance, using (\ref{conduct4}) then (punching $\phi=0$) gives
\begin{equation}\label{tan}
\sigma_{1,1} = \frac{1}{2\psi} \frac{e^2}{2\hbar \pi^2}\,\dashint_{-\pi}^{+\pi} d\phi\, \kappa^2(E_F)\tan(\phi)H(\phi+\psi)H(\psi-\phi)
\end{equation}
Or, equally
\begin{equation}\label{tan2}
\sigma_{1,1} = \frac{\kappa^2(E_F)}{2\psi} \frac{e^2}{2\hbar \pi^2}\,\dashint_{-\psi}^{+\psi} d\phi\, \tan(\phi)=-C\left[\log|\cos(\phi)|\right]_{-\psi}^{+\psi}
\end{equation}
with $C=\frac{\kappa^2(E_F)}{2\psi} \frac{e^2}{2\hbar \pi^2}$. From (\ref{tan2}) follows $\sigma_{1,1}=-C\{\log|\cos(\psi)|-\log|\cos(-\psi)|\}$. Hence, $\sigma_{1,1}=0$. A similar argument applies to $\sigma_{2,2}$ where the integration shows that $\sigma_{2,2}=-C\{\log|\sin(\psi)|-\log|\sin(-\psi)|\}=0$. If we subsequently turn to $\sigma_{1,2}=\sigma_{2,1}$ then we see
\begin{equation}\label{tancot}
\sigma_{1,2} = \frac{\kappa^2(E_F)}{2\psi} \frac{e^2}{2\hbar \pi^2}\,\dashint_{-\psi}^{+\psi} d\phi \sqrt{ \tan(\phi)}\sqrt{ \cot(\phi)}= \frac{e^2\kappa^2(E_F)}{2\hbar \pi^2}
\end{equation}
Hence, with $\kappa(E_F)=\frac{\pi\sqrt{2\hbar}}{e}$ the Pauli matrix $\sigma_x$ can be obtained for the conductance matrix. However, because of infinitesimal changes in 'associated velocity' (\ref{Paul1}), the factor is maintained. Hence,  
\begin{eqnarray}\label{Paulixlike}
\Sigma_x=   \frac{e^2\kappa^2(E_F)}{2\hbar \pi^2} \left( \begin{array}{cc} 0 & 1\\
1 & 0 \end{array}\right) 
\end{eqnarray}
$\kappa$ sufficiently small. Note that 'going through a singularity' for the $\cot$ integral provides a zero result of the integral when the $\phi=0$ is 'cut out' of the integration by left-hand $-\epsilon^{\prime}$ and right-hand $\epsilon^{\prime}$ for $\epsilon^{\prime} \rightarrow 0^{+}$. This is so because $|\sin(-\epsilon^{\prime})|$ equals $|\sin(\epsilon^{\prime})|$. Hence, the integral of both the $\tan$ as well as the $\cot$ containing associated 'velocities' in equation (\ref{Paul1}) can be given by the  $\epsilon^{\prime} \rightarrow 0^{+}$ sum of  $\int_{-\pi}^{-\epsilon^{\prime}}$ and $\int_{\epsilon^{\prime}}^{\pi}$ integration operations and is written $\dashint_{-\pi}^{+\pi}$. The Heaviside functions in (\ref{Paul1}) change the integrals into $\dashint_{-\psi}^{+\psi}$.

\subsubsection{$\sigma_y$ conductance}\label{P2}
The $\sigma_y$ Pauli matrix can be similarly derived for the conductance. Let us define the 'associated velocity' entries
\begin{eqnarray}\label{Paul2}
\begin{array}{cc} \delta u_{y,1}=u_1(E_F,\phi)=\lambda(E_F)\cos(\phi /2)H(\phi-\eta)H(\frac{\pi}{2}-\phi)\\
\delta u_{y,2}=u_2(E_F,\phi)=i\,\lambda(E_F)\sin(\phi /2)H(\phi-\frac{\pi}{2})H(\pi-\eta-\phi) \end{array}
\end{eqnarray}
with $\eta \in (0,\psi)$. In the first place we may note that, dissimilar to the $\sigma_x$ case in section - \ref{velvec}, the canceling of 'associated' kinetic energy terms occur in a non-symmetrical way: i.e. $\frac{1}{2}m^{*}||\mathbf{u}(\phi)||^2+ \frac{1}{2}m^{*}||\mathbf{u}(\phi^{\,\prime})||^2=0$ where $(\phi^{\,\prime},\phi)$ solves $\cos^2(\phi)-\sin^2(\phi^{\,\prime})=0$ with $\phi \in (\eta,\frac{\pi}{2})$ and $\phi^{\,\prime} \in (\frac{\pi}{2},\pi-\eta)$. If the aim is to explain the geometric asymmetry in electron - hole creation geometry then perhaps the crystal structure, i.e. phonon hindrance can be held accountable for the asymmetry. 
Subsequently, the integral for $u_1(E_F,\phi)u_2(E_F,\phi)$ vanishes because the domains are disjoint. For the $\sigma_{1,1}$ term we see
\begin{equation}\label{sigm211}
\sigma_{1,1}=\frac{e^2\lambda^2(E_F)}{2\hbar \pi^2}\int_{\eta}^{\pi /2} \cos^2(\phi /2) \,d\phi
\end{equation}
Because $\cos^2(\phi /2)=\frac{1}{2}(1+\cos(\phi))$ it follows that \\ $\sigma_{1,1}=\frac{e^2\lambda^2(E_F)}{4\hbar \pi^2}\left\{\frac{\pi}{2}-\eta+1-\sin(\eta)\right\}$. Similarly for $\sigma_{2,2}$ we may derive
\begin{equation}\label{sigm222}
\sigma_{2,2}=-\frac{e^2\lambda^2(E_F)}{2\hbar \pi^2}\int_{\pi /2}^{\pi-\eta} \sin^2(\phi /2) \,d\phi
\end{equation}
Because $\sin^2(\phi /2)=\frac{1}{2}(1-\cos(\phi))$ it follows that \\ $\sigma_{2,2}=-\frac{e^2\lambda^2(E_F)}{4\hbar \pi^2}\left\{\frac{\pi}{2}-\eta+1-\sin(\eta)\right\}$. Hence, when we take $\lambda(E_F)$ and $\kappa(E_F)$ such that 
\begin{equation}\label{gauge}
\kappa^2(E_F)=\frac{\lambda^2(E_F)}{2}\left\{\frac{\pi}{2}-\eta+1-\sin(\eta)\right\}
\end{equation}
then we find
\begin{eqnarray}\label{Pauliylike}
\Sigma_y=   \frac{e^2\kappa^2(E_F)}{2\hbar \pi^2} \left( \begin{array}{cc} 1 & \,\,0\\
0 & -1 \end{array}\right) 
\end{eqnarray}

\section{QUANTUM CORRELATION}
The Pauli matrices can be derived from the Boltzmann treatment of a two dimensional electron gas generated from two different crystal structures at 90 degrees angle. The locally created electron gas in A and B wing of a Bell-type experiment enables the simulation of entangled particle pair correlation with classical local constructed currents. We will argue for a micro Boltzmann distribution interpretation of the EPR paradox. 
 
In order to have a simpler representation for the conductances let us assume that the (2D) electric field vector equals: $\mathbf{E}= \frac{2\hbar \pi^2}{e^2\kappa^2(E_F)} \hat{\mathbf{E}}$, with norm, $ ||\hat{\mathbf{E}}||^{2}= \hat{\mathbf{E}}^{T}\, \hat{\mathbf{E}}=1$. Hence, if $f_1\left(\Sigma_x , \Sigma_y\right)$ is a linear function then another linear function $f_2(\sigma_x,\sigma_y)$ exists such that a current $\mathbf{j}$ can be written as $\mathbf{j}=f_1\left(\Sigma_x , \Sigma_y\right)\mathbf{E}=f_2\left(\sigma_x , \sigma_y\right)\hat{\mathbf{E}}$. The 'particles' flying from the source to the instruments initiate the $\hat{\mathbf{E}}$ on both the $A$ and $B$ side instruments. Induced 2D currents in both the $A$ and $B$ wing explain the correlation.   

%\subsection{Entangled currents mixing algebra}
If we take the two 'crystal structures' and generate mixed conductance 2D electron gas, then at first instance, electrons in the 2D gas from the horizontal, $\Sigma_x$ related, crystal and from the vertical, $\Sigma_y$ related, crystal co-occur in the angular interval $(\eta,\psi)$. In this area per wing one can 'mix the two conductances' $\Sigma_x$ and $\Sigma_y$ using a unit parameter vector $\mathbf{\hat{a}}=(a_1,a_2)$ on the $A$ side and/or $\mathbf{\hat{b}}=(b_1,b_2)$ on the $B$. E.g.
%\begin{equation}\label{mixing}
$\Sigma_A(a_1,a_2)=a_1\Sigma_x + a_2\Sigma_y$. 
%\end{equation}
The mixed current at $A$ then is equal to $\mathbf{j}_A=\Sigma_A(a_1,a_2)\mathbf{E}$. Similarly, one can define a $\Sigma_B(b_1,b_2)$ such that $\mathbf{j}_B=\Sigma_B(b_1,b_2)\mathbf{E}$. 
 
We can take $f(a_1) \times 100\%$ to indicate the percentage at the $A$ wing of $\Sigma_x$-crystal electrons and similarly for the $g(a_2) \times 100\%$. The same thing can be supposed for the entries of the $\mathbf{b}$ vector at the B side of the experiment.
Furthermore, as an example the mixing percentages are transformed into angles. We suppose, $a_1=\cos(\theta_A)$ and $a_2=\sin(\theta_A)$ together with $b_1=\cos(\theta_B)$, $b_2=\sin(\theta_B)$ and project the $\theta $ in an interval $[\gamma, \tau] \subset (0,\frac{\pi}{2})$.  Subsequently, we may note that because of $\mathbf{E}^{T}=\frac{2\hbar \pi^2}{e^2\kappa^2(E_F)}(E_1,E_2)$ and the expressions for $\Sigma_x$ and $\Sigma_y$ in resp. (\ref{Paulixlike}) and (\ref{Pauliylike}) the following inner-product for $\mathbf{j}_A$ and $\mathbf{j}_B$ employs Pauli matrices as in a quantum corelation (see e.g. Bell\cite{Bell64}).
\begin{equation}\label{innerprd}
\mathbf{j}^{T}_A\,\mathbf{j}_B=\mathbf{\hat{E}}^{T}\left[\cos(\theta_A)\sigma_x +\sin(\theta_A)\sigma_y\right]\left[\cos(\theta_B)\sigma_x +\sin(\theta_B)\sigma_y\right]\mathbf{\hat{E}}
\end{equation}
Note that, $\sigma_x^{T}=\sigma_x$ and $\sigma_y^{T}=\sigma_y$ together with $\sigma_x^2=\sigma_y^2=1_{2\times2}$. If we in addition inspect e.g. the term $ \cos(\theta_A)\sin(\theta_B)\mathbf{\hat{E}}^{T}\sigma_x\sigma_y\mathbf{\hat{E}}$ it is noted that this equals
%\begin{eqnarray}\label{difinnerprd}
 $\cos(\theta_A)\sin(\theta_B)(E_1,E_2) \left( \begin{array}{cc} -E_2 \\ E_1 \end{array}\right)=0$
%\end{eqnarray} 
Hence, the inner product of the A and B wing current is 
%\begin{equation}\label{jAjB}
$\mathbf{j}^{T}_A\,\mathbf{j}_B=\cos(\theta_A-\theta_B)$.
%\end{equation}
The inner product of the two current vectors shows the entanglement but note it is build from local currents and conductances.
%\subsection{The experiment}
The previous 2D electron gas 'simulation of the quantum correlation' can be experimentally researched. Suppose, the source in a Bell-type experiment can be the cloning and subsequent sending in two different directions of the electric field vector $\mathbf{E}$. Two pairs of crystal structures are employed to generate the mixing of conductances $\Sigma_x$ and $\Sigma_y$, in the $A$ and conductances $\Sigma_x$ and $\Sigma_y$, in the $B$ wing. The electrons escape from the crystal surfaces ($\Sigma_x$ parallel $x^{+}$ axis and $\Sigma_y$ parallel $y^{+}$ axis and $x \bot y$)  and locally mix in $\phi \in (\eta,\psi)$. If the current vectors that are created from the 'cloned' electric fields vectors are transported to a measuring system $O$ a current-current entangled inner product  $\mathbf{j}^{T}_A\,\mathbf{j}_B=\mathbf{a}\cdot \mathbf{b}$ can be observed from $\mathbf{j}_A=[a_1\sigma_x+a_2\sigma_y]\mathbf{\hat{E}}_A$ and $\mathbf{j}_B=[b_1\sigma_x+ b_2\sigma_y]\mathbf{\hat{E}}_B$. Hence, $\mathbf{j}^{T}_A\,\mathbf{j}_B$ can in principle violate the CHSH but note that the current vectors are created by local means. The parameters $\mathbf{a}$ and $\mathbf{b}$ refer to mixing percentages of electrons with $\phi \in (\eta,\psi)$: i.e. in the $A$ wing $f(a_1) \times 100\%$ of the electrons from the $\Sigma_x$ crystal and $g(a_2) \times 100\%$ from the $\Sigma_y$ crystal and similarly in the B wing. The $f$ and $g$ functions represent the amount of electrons to obtain the weigths of the Pauli conductances. Perhaps one would like to argue against entanglement in currents but the outcome in the innerproduct  $\mathbf{j}^{T}_A\,\mathbf{j}_B$ is the same as quantum mechanically. It must be noted that entanglement refers to something unobservable and obtains its meaning from its use\cite{Wittgenstein}. Essentially entanglement is concluded from experimental observations. Unless we in the present local intrumentalist treatment refered unwittingly to macroscopic quantum electron gas, Bell's correlation needs not be nonlocal in its origin. It can be claimed that the structure laid down here is the physics of locality and causality that Einstein had in mind. The question then of course is: 'if referred to the microphysics domain, what does the distribution in the Boltzmann equation distribute'. A preliminary answer could be that the distribution governs electrons or, perhaps even better, charged anyons and that they only obtain their individuality in an ensemble and in this way a measurement is accomplished.

\section{MACH-ZEHNDER CONDUCTANCE ALGEBRA}
Subsequently,  the instruments participating in the Mach-Zehnder interferometer are associated to certain conductance forms and finally the possible chains of MZ conductance algebra is demonstrated to lead to the explanation of Young's double slit interference and Wheeler's backward causation. 
 
The elements of the Mach-Zehnder interferometer are Beamsplitters ($\Phi$), Mirrors ($M$), Phase Shift ($\Delta$) and measurement interaction. Let us with the algebra first associate current transformation forms to the $\Phi$, $M$ and $\Delta$. Measurement is taking an inner product $p_{12}=\mathbf{j}_{1}^{T}\cdot \mathbf{j}_{2}$ of currents. In our notation, 'in the index' of the symbol the sequence of optical elements is represented, e.g. $\mathbf{j}_{B_{\rightarrow}\,\Delta}$ means the current in the $B_{\rightarrow}$ path of the MZ interferometer 'after it passed' the phase shifter $\Delta$ which produced a shift in phase $\delta$. 
\subsection{The beam splitters $\Phi$}
The effect of a beamsplitter on the conductance description is that two branches of conductance arises. 
\begin{eqnarray}\label{BS}
\Phi\,:
\begin{array}{cc}
\sigma_{\,\rightarrow}=\frac{1}{\sqrt{2}}\left(\sigma_x - \sigma_y \right) \\
\sigma_{\,\uparrow}=\frac{1}{\sqrt{2}}\left(\sigma_x + \sigma_y \right) 
\end{array}
\end{eqnarray} 
Following the MZ discussion of Mittelstaedt \cite{Mittel} we have an entrance beamsplitter $\Phi_B$, with  the current related to $\sigma_{B_{\rightarrow}}$ parallel the entrance beam and the current related to $\sigma_{B_{\uparrow}}$ orthogonal the entrance beam. Moreover, there is a removable second beamsplitter $\Phi_A$, with conductances $\sigma_{A_{\rightarrow}}$ and $\sigma_{A\,\uparrow}$. The current related to $\sigma_{A_{\rightarrow}}$ is parallel the ${B_{\rightarrow}}$ path. The current related to $\sigma_{A\,\uparrow}$ is parallel the ${B_{\uparrow}}$ path.   
 
Note that the possible 2D currents derived from the conductance associated to e.g. beamsplitter $\Phi_B$ are $\mathbf{j}_{B_{\rightarrow}} = \sigma_{B_{\rightarrow}} \hat{\mathbf{E}}$ and $\mathbf{j}_{B_{\uparrow}} = \sigma_{B_{\uparrow}} \hat{\mathbf{E}}$. From equation (\ref{BS}) it follows that the inner product $\mathbf{j}_{B_{\rightarrow}}^{T}  \cdot \mathbf{j}_{B_{\uparrow}}=0$. This is so because $\hat{\mathbf{E}}^{T}\sigma_z\hat{\mathbf{E}}=0$, with, $\sigma_z=i\sigma_x \sigma_y$ . Hence, one route excludes the other. The "welcher Weg" (i.e. which path) question can be recognized in this orthogonality of currents. 
\subsection{The mirrors $M_i$}
Let us first consider the influence of $\sigma_z$ on the $\sigma_x$ and $\sigma_y$. We have, $\sigma_z\sigma_y=i\sigma_x$ and $\sigma_z\sigma_x=-i\sigma_y$. It then follows from the Pauli matrices that $\sigma_z \sigma_{B_{\rightarrow}} \sigma_z=-\sigma_{B_{\rightarrow}}$ and $\sigma_z \sigma_{B_{\uparrow}} \sigma_z=-\sigma_{B_{\uparrow}}$. Hence, $\sigma_{B_{\rightarrow}}=i\sigma_{B_{\uparrow}}\sigma_z$ and $\sigma_{B_{\uparrow}}=-i\sigma_{B_{\rightarrow}}\sigma_z$. The mirrors have the following (post multiplication) effect on the conductance in the MZ interferometer.
\begin{eqnarray}\label{Mir}
M\,:
\begin{array}{cc}
~~~M_2=M_{\rightarrow}=-i\sigma_z \\
M_1=M_{\uparrow}=i \sigma_z
\end{array}
\end{eqnarray}
The mirror $M_{\rightarrow}$ is in the path $B_{\rightarrow}$. The mirror $M_{\uparrow}$ is in the path $B_{\uparrow}$. Hence, from equation (\ref{Mir}) we see that the 2D conductances after passing the mirrors are either $\sigma=\sigma_{B_{\rightarrow}}M_2=\sigma_{B_{\rightarrow}}M_{\rightarrow}=i\sigma_{B_{\uparrow}}\sigma_z(-i)\sigma_z=\sigma_{B_{\uparrow}}$, passing $M_2=M_{\rightarrow}$, or  $\sigma=\sigma_{B_{\uparrow}}M_1=\sigma_{B_{\uparrow}}M_{\uparrow}=-i\sigma_{B_{\rightarrow}}\sigma_z\,i\,\sigma_z=\sigma_{B_{\rightarrow}}$, passing $M_1=M_{\uparrow}$. Hence the mirrors do not affect the orthogonality of the two currents i.e. the routes remain mutually exclusive. The currents conductances changes. In the $B_{\uparrow}$ path after the mirror $M_{\uparrow}$ the conductance is changed from $\sigma_{B_{\uparrow}}$ to $\sigma_{B_{\rightarrow}}$. Similarly a change in the $B_{\rightarrow}$ occurs after the mirror, $M_{\rightarrow}$, with $\sigma_{B_{\rightarrow}}$ transforming to $\sigma_{B_{\uparrow}}$. Note that if a beam splitter $\Phi_C$ is put in place of $M_{\uparrow}$ the $j_{C_{\uparrow}}$ current would be parallel the $B_{\uparrow}$ path. Hence the $j_{C_{\rightarrow}}$ current would be measured against the $j_{B_{\rightarrow}\,M_{\rightarrow}}$ current, i.e the current in the beam that met mirror $M_{\rightarrow}$. This would still be mutual exclusive currents: $P=j_{B_{\rightarrow}\,M_{\rightarrow}}^{T} \cdot j_{C_{\rightarrow}}=0 $.  
\subsection{Phase shifter $\Delta$ and the complete MZ configuration}
Let us suppose that the phase shift is $\delta$ size. We suppose that the phase shift affects the $B_{\rightarrow}$ branch. In the algebra we take the $\sigma_y$ part of the conductance to be affcted by the $\cos(\delta)$ influence. We have
\begin{equation}\label{Delta}
\sigma_{B_{\rightarrow}}=i\sigma_{B_{\uparrow}}\sigma_z~ \,\xrightarrow{\Delta}\, ~\frac{i}{\sqrt{2}}\left(\sigma_x + \cos(\delta)\sigma_y\right)\sigma_z=i\sigma_{B_{\uparrow}}(\delta)\sigma_z
\end{equation} 
Note that the transformation with $\cos(\delta)$ could also have been incorporated thus 
\begin{equation}\label{Deltato}
\sigma_{B_{\rightarrow}}\, \xrightarrow{\Delta}\,\frac{1}{\sqrt{2}}\left(\sigma_x - \cos(\delta)\sigma_y\right)=\sigma_{B_{\rightarrow}}(\delta)
\end{equation}
and after the $\Delta$ transformation we have $\sigma_{B_{\rightarrow}}(\delta)=i\sigma_{B_{\uparrow}}(\delta)\sigma_z$ etc.
 
If the $B_{\rightarrow}$ route contains $\Delta$ and $M_{\rightarrow}$, the notation for the influence of all those elements is e.g. for the conductance: $\sigma_{B_{\rightarrow} \,\Delta\,M_{\rightarrow}}$. If this conductance results in a current vector we obtain
\begin{equation}\label{chain} 
\mathbf{j}_{B_{\rightarrow} \,\Delta\,M_{\rightarrow}}=\sigma_{B_{\rightarrow} \,\Delta\,M_{\rightarrow}}\hat{\mathbf{E}}
\end{equation}
Because of the definition of the elements in the MZ interferometer it is easy to see that the 2D conductance in the $\rightarrow$ path is equal to
\begin{equation}\label{Chain2}
\sigma_{B_{\rightarrow} \,\Delta\,M_{\rightarrow}}=\frac{1}{\sqrt{2}} \left( \sigma_x + \cos(\delta) \sigma_y\right)
\end{equation}
Hence, the quantum result $P(\varphi;A_{\rightarrow})$ can be written as $\mathbf{j}_{B_{\rightarrow} \,\Delta\,M_{\rightarrow}}^{T} \cdot \mathbf{j}_{A\, \rightarrow}$
because
\begin{equation}\label{inner1}
\mathbf{j}_{B_{\rightarrow} \,\Delta\,M_{\rightarrow}}^{T} \cdot \mathbf{j}_{A_{\rightarrow}}=\frac{1}{2}\hat{\mathbf{E}}^{T}\left(\sigma_x + \cos(\delta)\sigma_y\right)\left(\sigma_x - \sigma_y\right)\hat{\mathbf{E}}
\end{equation}
From the previous equation we then derive that 
\begin{equation}\label{Inner2}
\mathbf{j}_{B_{\rightarrow} \,\Delta\,M_{\rightarrow}}^{T} \cdot \mathbf{j}_{A_{\rightarrow}}=\frac{1}{2}\left(1-\cos(\delta)\right)
\end{equation}
such that with some elementary mathematics: $\mathbf{j}_{B_{\rightarrow} \,\Delta\,M_{\rightarrow}}^{T} \cdot \mathbf{j}_{A_{\rightarrow}}=\sin^2(\frac{\delta}{2})$. This is the value predicted by quantum mechanics for the detector in the outgoing line of mirror $M_{\rightarrow}$, see Mittelstaedt \cite{Mittel}. Similarly we can with the MZ conductance algebra derive that $\mathbf{j}_{B_{\rightarrow} \,\Delta\,M_{\rightarrow}}^{T} \cdot \mathbf{j}_{A_{\uparrow}}$ is the value that is predicted by quantum mechanics, $P(\varphi;A_{\uparrow})$, for the detector in the outgoing line of $M_{\uparrow}$. Hence,
\begin{equation}\label{inner3}
\mathbf{j}_{B_{\rightarrow} \,\Delta\,M_{\rightarrow}}^{T} \cdot \mathbf{j}_{A_{\uparrow}}=\frac{1}{2}\hat{\mathbf{E}}^{T}\left(\sigma_x + \cos(\delta)\sigma_y\right)\left(\sigma_x + \sigma_y\right)\hat{\mathbf{E}}=\frac{1}{2}\left(1+\cos(\delta)\right)
\end{equation}
Hence, $\mathbf{j}_{B_{\rightarrow} \,\Delta\,M_{\rightarrow}}^{T} \cdot \mathbf{j}_{A_{\uparrow}}=\cos^2(\frac{\delta}{2})$, see Mittelstaedt \cite{Mittel}.
 
At the end of this section it is noted that a reduction rule is necessary in the algebra to deal with equivalent forms of extended MZ configurations. E.g. if there are two phase shifters in the $B_{\rightarrow}$ path then they are treated as a single phase shifter where the $\delta$'s are summed. Moreover, Mirror and phase shifter can be interchanged without affecting the result. If so, then in the algebra e.g. the $\sigma_{B_{\rightarrow} M_{\rightarrow} \Delta}$ equals $\sigma_{B_{\rightarrow}\Delta M_{\rightarrow}}$. 
%\subsection{Current in A}
This leaves us with one final point that is explained below. What creates the $\Phi_A$ current. Let us answer this question by making a division between "pump" and "signal" current. In the first place e.g. a photon $\lambda$ enters the beamsplitter. We assume it carries a "pump" 2D current $\mathbf{j}_{\lambda}=\sigma_{\lambda}\hat{\mathbf{E}}$. We can e.g. assume that $\sigma_{\lambda}=1_{2\times2}$. At this point the nature of $\sigma_{\lambda}$ is not so very important. As we discussed in the paper two currents arise, $\mathbf{j}_{B_{\uparrow}}$  and $\mathbf{j}_{B_{\rightarrow}}$. Let us assume the $\mathbf{j}_{B_{\rightarrow}}$ is parallel to the entrance beam $\mathbf{j}_{\lambda}$. Because the phase shifter $\Delta$ is in the $B_{\rightarrow}$ beam and that beam is parallel to the entrance beam, we call that the "signal" beam and $\mathbf{j}_{B_{\rightarrow}}$ the "signal" current. The current $\mathbf{j}_{B_{\uparrow}}$ is called the "pump" current. Now, $\mathbf{j}_{B_{\uparrow}}$ meets the $M_{\uparrow}$ mirror and transforms to "pump current" $\mathbf{j}_{B_{\uparrow}\,M_{\uparrow}}=\sigma_{B_{\rightarrow}}\hat{\mathbf{E}}$. At "the same time"   the signal current $\mathbf{j}_{B_{\rightarrow}}$ meets the phase shifter $\Delta$ and then the mirror $M_{\rightarrow}$ and transforms into $\mathbf{j}_{B_{\rightarrow}\,\Delta\, M_{\rightarrow}}=\sigma_{B_{\uparrow}}(\delta)\hat{\mathbf{E}}$. Now, in case of no $\Phi_A$, from the conductances in both "pump" and "signal" one can easily see that for $\delta=0$ the "welcher Weg" (which route) question can be answered because $\hat{\mathbf{E}}^{T}\sigma_{B_{\rightarrow}}\sigma_{B_{\uparrow}}\hat{\mathbf{E}}=0$ and the "pump" and ($\delta=0$) "signal" currents are mutual exclusive. Depending on the preparation state of the measuring instruments: if the "pump" current is picked up first, we say the photon went through the $B_{\uparrow}$ branch. If the "signal current is picked up first we say the photon went through the $B_{\rightarrow}$ route. Note the "opposite current" is nullified in the measurement. For completeness we may note the introduction of a $\Phi_C$ instead of mirror $M_{\uparrow}$. The orthogonality is maintained irrespective a mirror or beamsplitter is used to obtain the 'horizontal' $B_{\uparrow}$ beam. 

Suppose the $\Phi_A$ is present and we call $\mathbf{j}_{B_{\uparrow}}$ "pump" then we may say that the $\mathbf{j}_{A_{\rightarrow}}$ and $\mathbf{j}_{A_{\uparrow}}$ are pumped by $\mathbf{j}_{B_{\rightarrow}}$ and that measurement, e.g. with $\delta \neq 0$  is the inner product of $\mathbf{j}_{B_{\rightarrow}\,\Delta\, M_{\rightarrow}}$ with $\mathbf{j}_{A_{\rightarrow}}$ and $\mathbf{j}_{A_{\uparrow}}$. In the present paper it was demonstrated that the innerproducts: $\left( \mathbf{j}_{B_{\rightarrow}\,\Delta\, M_{\rightarrow}}^{T} \cdot \mathbf{j}_{A_{\rightarrow}}\right)=\cos^2(\delta/2)$ and $\left( \mathbf{j}_{B_{\rightarrow}\,\Delta\, M_{\rightarrow}}^{T} \cdot \mathbf{j}_{A_{\uparrow}}\right)=\sin^2(\delta/2)$ correspond to the values predicted by quantum mechanics.    

Concludingly, with the MZ conduction algebra the quantum interference effect plus phase shifting is explained with instrument based changes in the conductance. The idea is that the particle (photon) is related to the simple $\hat{\mathbf{E}}$ vector that causes the current. In the instrument ensembles of charged particles follow Boltzmann's transport equation and underpin the measurement. They in fact only 'exist' in the ensemble. The conclusion from the previous paragraphs is that the insertion of the beam splitter $\Phi_A$ produces the interference effect because the 2D conductance of the final state of the measurement set-up changes. 

\section{CONCLUSION}
It was demonstrated that Pauli matrices can be obtained in 2D conductance tensors. Subsequently it was demonstrated that the Pauli conductance can be manipulated such that the quantum correlation arises. An experiment was described to simulate the quantum correlation in 'macroscopic' reality. Subsequently, the Pauli conductance was applied to one-photon Mach-Zehnder interferometry. Each element in the interferometer was associated to a transformation of the 2D conductances, hence, of the current in 2D. This gave a MZ conductance algebra. Elements in the interferometer were identified together with rules how to transform the propagating current. Finally it was demonstrated that the quantum values were obtained from the inner product of the transformed current and the second beamsplitter currents. 

When the $\Phi_A$ is absent there is a different conduction current $\mathbf{j}$ compared to the presence of $\Phi_A$. Wheeler's backward causation when the splitter $\Phi_A$ is inserted after the particle has passed the first splitter $\Phi_B$ can be explained by man-made activities related to the {\it configuration of the measuremental setup}. Similarly the particle-wave duality is explained because in the MZ interferometer the "welcher Weg" question is answered without $\Phi_A$. Self-interference supposedly occurs when $\Phi_A$ is inserted but is expressed in terms of MZ algebra.
 
Hence, quantum paradoxes are explained by unknown influences of the instruments and their configuration in the experiment. However, if a photon is seen as a response of instruments, then what are the 2D electrons or charged anyons used in the explanation. This leads us to acknowledge that quantum theory is a warning that otherwise meaningful concepts may loose meaning \cite{Wittgenstein} in quantum reality. The meaning-of-language border aims to differentiate metaphysical concepts from concepts without observational unclarity.  Because of the demonstrated influence of the measurement instrument one can say that a different, more man-made, view of the duality particle-wave is obtained. The price to pay is that particles only exist in ensembles. This postulate can be studied in experiment however. A similar concept in physical theory is the impossibility that quarks exist as separate particles.
 
The idea of individuation in an ensemble itself may be meaningless. This also refers to the impossibility to formulate concepts about quantum phenomena without refering to instruments. If the concepts of instrumentalism loose their meaning, the only thing that can be said in favour of it is that it crossed the border at a new point. Even interpreting a click in a measuring  instrument as: 'a quantum particle just hit the detector' is stricly speaking not free from idealistic metaphysical contamination. It appears beneficial to note that in foundation science one form of metaphysics is likely to be replaced by an other, not necessarily better, but different form. In this sense pointing at the influence of measurement instruments in paradox points at man-made parts of quantum reality.

\end{document}